\documentclass[
journal
,onecolumn
]{IEEEtran}
\usepackage{amsmath}
\usepackage{amssymb}
\usepackage{amsfonts}
\usepackage{amsthm}
\usepackage{subfigure}
\usepackage{color}
\usepackage{graphicx}
\usepackage{array}
\usepackage{tikz}
\usepackage{cite}
\usepackage{verbatim}
\usepackage{algorithm}
\usepackage{algpseudocode}
\usepackage[center]{caption}
\newtheorem{theorem}{Theorem}
\newtheorem{lemma}{Lemma}
\newtheorem{corollary}{Corollary}
\newtheorem{mydef}{Definition}

\begin{document}

\newcommand{\blockcomment}[1]{}

\title{Two Approaches to the Construction of Deletion Correcting Codes: Weight Partitioning and Optimal Colorings}
\author{Daniel~Cullina,~\IEEEmembership{Student Member,~IEEE}
        Ankur~A.~Kulkarni,
        and Negar~Kiyavash,~\IEEEmembership{Member,~IEEE}%
\thanks{The material in this paper was presented (in part) at the Internation Symposium on Information Theory, Cambridge, MA, USA, July 2012 \cite{cullina_coloring_2012}.
This work was supported in part by AFOSR under grants FA 9550-11-1-0016 and FA 9550-10-1-0573; and by NSF grants CCF 10-54937 CAR and CCF 10-65022 Kiyavash.}%
\thanks{Daniel Cullina is with the Department of Electrical and Computer Engineering and the Coordinated Science Laboratory, University of Illinois at Champaign-Urbana, Urbana, Illinois 61801 (email: cullina@illinois.edu). }%
\thanks{Ankur Kulkarni is with the Coordinated Science Laboratory, University of Illinois at Champaign-Urbana, Urbana, Illinois 61801 (email: akulkar3@illinois.edu). }%
\thanks{Negar Kiyavash is with the Department of Industrial and Enterprise Systems Engineering and the Coordinated Science Laboratory, University of Illinois at Champaign-Urbana, Urbana, Illinois 61801 (email: kiyavash@illinois.edu). }%
}
\maketitle
\begin{abstract}
We consider the problem of constructing deletion correcting codes over a binary alphabet and take a graph theoretic view.
An $n$-bit $s$-deletion correcting code is an independent set in a particular graph.
We propose constructing such a code by taking the union of many constant Hamming weight codes.
This results in codes that have additional structure.
Searching for codes in constant Hamming weight induced subgraphs is computationally easier than searching the original graph.
We prove a lower bound on size of a codebook constructed this way for any number of deletions and show that it is only a small factor below the corresponding lower bound on unrestricted codes.
In the single deletion case, we find optimal colorings of the constant Hamming weight induced subgraphs.
We show that the resulting code is asymptotically optimal.
We discuss the relationship between codes and colorings and observe that the VT codes are optimal in a coloring sense.
We prove a new lower bound on the chromatic number of the deletion channel graphs.
Colorings of the deletion channel graphs that match this bound do not necessarily produce asymptotically optimal codes.
\end{abstract}

\section{Introduction}
\IEEEPARstart{D}{eletion} channels output only a subsequence of their input while preserving the order of the transmitted symbols. 
They have applications in biology, synchronization problems, and communication of information over packet networks.
This paper concerns channels that take a binary input string of fixed length and a delete a fixed number of symbols. 
Despite significant effort on this case, there still are many fundamental open problems.
In particular, we are interested in the design of $s$-deletion correcting codes and the cardinality of the largest possible codebook.

Levenshtein gave partial answers to both problems.
He derived asymptotic upper and lower bounds on the sizes of codes for any number of deletions \cite{levenshtein_binary_1966}.
He showed that the Varshamov Tenengolts (VT) codes, which had been designed to correct a single asymmetric error \cite{varshamov_codes_1965,varshamov_arithmetic_1965}, could be used to correct a single deletion.
The VT codes meet the upper bound, so they are asymptotically optimal and establish the capacity of the single deletion channel.

This paper addresses two questions related to code construction by taking a graph theoretic perspective.
For each input string length and number of deletions, there is a graph that expresses all of the constraints on code construction.
The vertices of this graph correspond to the binary strings of that length and a code is an independent set in the graph.
The problem of finding a maximum independent set is NP Hard for general graphs.

First, we present a two stage method for code construction.
The method involves partitioning the vertices of the graph according to Hamming weight, finding codes in selected partitions, and taking the union of these codes.
The substrings of a particular weight form a subgraph. 
Independent sets in this subgraph can be found in various ways, in particular, by exhaustive search, greedy search, or explicit graph coloring.
Finding good codes in the subgraphs is less computationally intensive than exhaustively searching within the whole graph.
For any number of deletions we prove a lower bound on the size of codes constructed using these subgraphs.
This bound is within a small constant factor of the Levenshtein lower bound.
This demonstrates that adding this restriction on codeword weights requires us to pay only a small penalty in the code sizes that we can guarantee.
In the single deletion case, we use this method to construct new asymptotically optimal codes.
These use an optimal coloring of the constant weight subgraphs.

Second, having taken graph theoretic perspective, we ask if the existing codes of Varshamov and Tenengolts have a graph interpretation.
We observe that VT codes are optimal colorings of the whole single deletion graphs.
Any sequence of optimal colorings of the single deletion graphs produces sequences of codes that match Levenshtein's upper bound.
We show that the same is not true for the multiple deletion graphs by deriving a lower bound on the chromatic number of the graphs for each string length and number of deletions.
Even if there are sequences of colorings using the number of colors specified by the bound, the coresponding sequences of codes are not guaranteed to match the Levenshtein upper bound.
Consequently, either solving the coloring problem for multiple deletions is not sufficient for finding asymtotically optimal independent sets, or Levenshtein's upper bound on independent set size is not tight.

\subsection{Related Work}
A wide variety of code constructions have been proposed for the deletion channel and other closely related channels.
These constructions vary significantly in code size, explicitness, and efficiency of construction so all comparisons must be done carefully.
Tenengolts found an asymptotic upper bound on single deletion correcting codes over nonbinary alphabets.
He constructed codes over each $q$-ary alphabet that are within a factor of $\frac{q}{q-1}$ of the bound \cite{tenengolts_nonbinary_1984}.
Helberg and Ferreira attempted to generalize the VT construction to any number of deletions, but the size of the resulting codes are far below Levenshtein's lower bound \cite{helberg_multiple_2002}.
Schulman and Zuckerman considered a different asymptotic regime.
They constructed nonexplicit but efficiently constructable codes for a channel that deletes a constant fraction of the symbols in each block \cite{schulman_asymptotically_1999}.

Another direction for the construction of codes is computational.
It is well known that the problem of finding deletion correcting codes is equivalent to finding an independent set in a particular graph \cite{sloane_single-deletion-correcting_2002}.
But since, for general graphs, finding the maximum independent set is NP-hard, exact algorithms rapidly become intractable with increasing input string length ($n$). 
Codes found via search usually lack structure and efficient decoding algorithms, but they are still interesting because they establish lower bounds on the size of optimal codes.
For the case of the single deletion, the computational approach has established that VT codes are optimal for $n\leq 10$ (graph with $2^{10}$ vertices)~\cite{sloane_challenge_????}.  
For multiple deletions, the best known codes have all been found through search algorithms. 
Butenko et al. found two-deletion correcting codes of maximum size for $n \leq 10$ \cite{butenko_finding_2002}.
Khajouei et al. used a heuristic algorithm to find the largest known two deletion correcting codes for $n \leq 25$ \cite{khajouei_algorithmic_2011}.

There has been much work on constructions, which provide lower bounds, but progess on upper bounds has been rare.
Levenshtein eventually refined his original asymptotic bound (and the parallel nonbinary bound of Tenengolts) into a nonasymptotic version \cite{levenshtein_bounds_2002}.
Kulkarni and Kiyavash recently proved a better upper bound for an arbitrary number of deletions and any alphabet size \cite{kulkarni_non-asymptotic_????}.

There are several other lines of work attacking related combinatorial problems.
One of these involves characterizing the sets of superstrings and substrings of any string.
Levenshtein showed that the number of superstrings does not depend on the starting string \cite{levenshtein_elements_1974}.
Calabi and Hartnett gave a tight bound on the number of substrings of each length \cite{calabi_general_1969}.
Hirschberg extended the bound to larger alphabets \cite{hirschberg_bounds_1999}.
Swart and Ferreira gave a formula for the number of distinct substrings produced by two deletions for any starting string \cite{swart_note_2003}.
Liron and Langberg improved and unified existing bounds and constructed tightness examples \cite{liron_characterization_2012}.

\subsection{Organization}
The paper is organized as follows. 
In Section~\ref{section:preliminaries}, we give some notation and definitions related to the deletion channel and review the graph theoretic terminology and results.
In Section~\ref{section:construction} we describe our code construction strategy and prove lower bounds on the sizes of the codes for any number of deletions. 
In Section~\ref{section:single-deletion} we construct new asymptotically optimal single deletion correcting codes and show that colorings used in the VT codes and in our codes are both optimal.
In Section~\ref{section:coloring} we discuss the relationship between optimal colorings and optimal independent sets for multiple deletion graphs and prove a lower bound on the number of colors needed for these graphs.
Proofs of some technical results are found in two appendices.
In Appendix~\ref{section:counting-superstrings}, we compute the weight distribution of the superstrings of a given string.
In Appendix~\ref{section:induced}, we identify various induced subgraphs in these graphs, demonstrating that the graphs are not perfect.

\section{Preliminaries}
\label{section:preliminaries}
\subsection{Notation}
Let $[n]$ be the set of nonnegative integers less than $n$, $\{0,1..n-1\}$.
Let $[2]^n$ be the set of binary strings of length $n$. 
Let $[2]^n_k$ be the set of binary strings of length $n$ with exactly $k$ ones.
Let $H(x)$ be the Hamming weight of a string $x$. 
We will need the following asymptotic notation: 
let $a(n) \sim b(n)$ denote that $\lim_{n \to \infty} \frac{a(n)}{b(n)} = 1$
and $a(n) \lesssim b(n)$ denote that $\lim_{n \to \infty} \frac{a(n)}{b(n)} \leq 1$.

We will use the following asymptotic equality frequently: for fixed $c$, $\binom{n}{c} \sim \frac{n^c}{c!}$.

\subsection{The deletion channel and associated graphs}
We will formalize the problem of correcting deletions by defining the deletion channel.
The deletion channel takes a binary string of length $n$ and outputs a substring of length $n-s$.
For binary strings $x$ and $y$, write $x<y$ if $x$ is a substring of $y$ and define the following sets.
\begin{mydef}
For $x \in [2]^n$, define 
\begin{equation*}
D_s(x) = \{z~\in~[2]^{n-s} | z<x\}, 
\end{equation*}
the set of substrings of $x$ that can be produced by $s$ deletions.
Similarly 
\begin{equation*}
I_s(x) = \{w \in [2]^{n+s} | w>x \},
\end{equation*}
the set of superstrings of $x$ that can be produced by $s$ insertions.
\end{mydef}
If $x$ is the input to an $n$ bit $s$ deletion channel, $D_s(x)$ is the set of possible outputs. 
If $x$ is the \textit{output} from the channel, $I_s(x)$ is the set of possible inputs.

When two inputs share common outputs they can potentially be confused by the receiver.
\begin{mydef}
For any two strings $x,y \in [2]^n$, define 
\begin{equation*}
D_s(x,y) = D_s(x) \cap D_s(y),
\end{equation*}
the set of common substrings of length $n-s$.
For any $x \in [2]^n$, define 
\begin{equation*}
N_s(x) = \{y \in [2]^n \setminus x | D_s(x,y) \neq \varnothing \},
\end{equation*}
the set of strings that share a common substring of length $n-s$ with $x$.
\end{mydef}

We are interested in codes that allow the correction of $s$ deletions.
\begin{mydef}
A length $n$ $s$-deletion correcting code is a set $C \subset [2]^n$ such that for any two distinct binary strings $x,y \in C$, $D_s(x,y)$ is empty.
A length $n$ $s$-deletion correction code is optimal if no larger code exists for those parameters.
A sequence of $s$-deletion correction codes with increasing $n$ is asymptotically optimal if the sequence of ratios of the their sizes to the optimal sizes goes to one.
\end{mydef}

We can also characterize codes by defining a distance measure on binary strings.
\begin{mydef}
Let $x \in [2]^m$ and $y \in [2]^n$ and let $z \in [2]^l$ be a common substring of $x$ and $y$ of maximum length.
Then $x$ can be transformed into $z$ by $m-l$ deletion operations and $z$ can be transformed into $y$ by $n-l$ insertion operations.
Thus the deletion distance between $x$ and $y$ is $d_L(x,y) = m+n-2l$.
\end{mydef}
It is well known that deletion distance is a metric \cite{levenshtein_binary_1966}.
If $x$ and $y$ are the same length, then the deletion distance between them is even and 
\begin{equation*}
d_L(x,y)/2 = \min \{s \in \mathbb{N} | D_s(x,y) \neq \varnothing \}.
\end{equation*}
Now we have a metric characteriztion of an $s$-deletion correcting code:
a set of codewords of length $n$ in which the deletion distance between any two codewords is greater than $2s$. 
Two codewords cannot both appear in a code if their deletion distance is $2s$ or less. We capture this condition by
defining the following graph.
\begin{mydef}
For all $s,n \in \mathbb{N}$, let $L_{s,n}$ be a graph 
with $[2]^n$ as its vertices. Vertices $x$ and $y$ are adjacent if and only if $d_L(x,y)/2 \leq s$.
\end{mydef}
Finally, we have a graphical characterization of an $s$-deletion correcting code:
a set of vertices in $L_{s,n}$ that have no edges between them.

\subsection{Independent Sets, Colorings, and Cliques}
\label{subsection:graph}
Now we will briefly define some graph notation and review a few concepts that will be useful later.
All of these are sourced from West \cite{west_introduction_2001}.
Given a graph $G$, let $V(G)$ denote its vertex set and let $E(G)$ denote its edge set.
Given $S \subseteq V(G)$, the \textit{subgraph induced by} $S$ contains the vertices in $S$ and the edges in $E(G)$ that have both endpoints in $S$.

An \textit{independent set} in a graph is a set of vertices 
that are all nonadjacent.
The size of a largest independent set in a graph $G$ is denoted by $\alpha(G)$.
The \textit{neighborhood} of a vertex is the set of adjacent vertices.
The \textit{degree} of a vertex is the number of adjacent vertices.
The maximum degree of any vertex in $G$ is denoted by $\Delta(G)$. 
Every maximal independent set contains at least $|V(G)|/(\Delta(G)+1)$ vertices.
This is because the union of the neighborhoods of the vertices in the independent set must contain all of the vertices in the graph.
The average degree of the vertices of $G$ is denoted by $\overline{d}(G)$.
Because each edge contributes to the degree of two vertices, $\overline{d}(G) = 2|E(G)|/|V(G)|$.
Some independent set containing at least $|V(G)|/(\overline{d}(G)+1)$ vertices always exists \cite[p. 122]{west_introduction_2001}.
This result is a version of Turan's Theorem.

A \textit{$k$-coloring} of a graph assigns a color (a element of $[k]$) to each vertex.
The coloring is proper if it never assigns the same color to both endpoints of an edge.
Thus a proper coloring of a graph partitions its vertices into independent sets; 
each independent set is assigned a single color and called a \textit{color class}.
The chromatic number of a graph $G$, denoted $\chi(G)$, is the smallest $k$ for which a proper $k$-coloring of $G$ exists.
An argument based on greedy coloring of $G$ shows that $\chi(G) \leq \Delta(G) + 1$.

A coloring gives us several independent sets to choose from, each corresponding to a color class.
At least one of these color classes must be at least as large as the average size of a color class.
Consequently, $\alpha(G)~\geq~|V(G)|/\chi(G)$. 
However, properly coloring a graph using the minimum number of colors is not equivalent to finding the largest independent set.
In general there is no guarantee that the largest color class in a particular coloring is a maximum independent set or that
any optimal coloring has a maximum independent set as a color class.

A \textit{clique} in a graph is a set of vertices that are all adjacent. 
The size of a largest clique in a graph $G$ is denoted by $\omega(G)$.
In a proper coloring, each vertex in a clique must be assigned a different color, so for any graph $G$, $\chi(G) \geq \omega(G)$.

For any graph $G$, we can define its $t$th power, denoted $G^t$.
The vertex sets of $G$ and $G^t$ are the same.
Vertices are adjacent in $G^t$ if and only if there is a path between them in $G$ of $t$ or fewer edges.
The neighborhood of any vertex in $G$ is a clique in $G^2$, so $\omega(G^2) \geq \Delta(G) + 1$.

Deletion distance satisfies the triangle inequality, so the length of the shortest paths between vertices $x$ and $y$ in $L_{s,n}$ is at most $d_L(x,y)/2s$.
This implies that if $x$ and $y$ are adjacent in $(L_{s,n})^t$, then $d_L(x,y)/2 \leq ts$.
Thus every edge in $(L_{s,n})^t$ is present in $L_{ts,n}$ and we have $\omega(L_{2s,n}) \geq \omega((L_{s,n})^2) \geq \Delta(L_{s,n}) + 1$.

\begin{figure}
\centering
 \begin{tikzpicture}[auto]
 \node (alpha)  at (0, 2)     {$\frac{|V(G)|}{\alpha(G)}$};
 \node (omega)  at (0, 1)     {$\omega(G)$};
 \node (chi)    at (2.5, 1.5) {$\chi(G)$};
 \node (ave)    at (2.5, 2.5) {$\overline{d}(G)+1$};
 \node (max)    at (5, 2)     {$\Delta(G)+1$};
 \node (omega2) at (7.5, 2)   {$\omega(G^2)$};

 \draw[->] (alpha) -- (ave);
 \draw[->] (alpha) -- (chi);
 \draw[->] (omega) -- (chi);
 \draw[->] (ave) -- (max);
 \draw[->] (chi) -- (max);
 \draw[->] (max) -- (omega2);
 \draw[->] (2.5,0.5) to node {$\leq$} (5,0.5);
 \end{tikzpicture}
\caption{Inequalities between graph parameters. 
}
\label{figure:inequalities}
\end{figure}
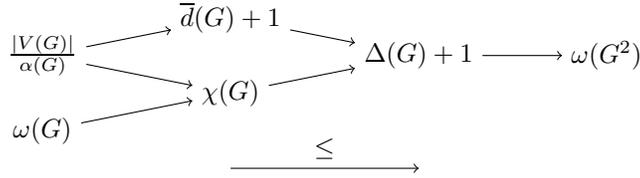

These inequalities are summarized in Fig.~\ref{figure:inequalities}.

\subsection{Existing results}
\label{subsection:existing}
Now that we have established some terminology and notation, we can concisely express some important existing results.
Levenshtein proved the following asymptotic upper and lower bounds on the size of optimal $s$-deletion correcting codes \cite{levenshtein_binary_1966}:
\begin{equation}
\label{eq:levenshtein-bounds}
\frac{2^{n+s}}{\binom{n}{s}^2} \lesssim \alpha(L_{s,n}) \lesssim \frac{2^n}{\binom{n}{s}}.
\end{equation}
We give a proof of the lower bound in Section~\ref{subsection:lower-bounds}.
Notice that there is a gap between the upper and lower bounds for all numbers of deletions.

For a single deletion, the VT construction asymptotically matches the upper bound and closes the gap.
The VT construction uses a weight function to partition $[2]^n$ into $n+1$ sets.
Levenshtein showed that each of these sets is a code \cite{levenshtein_binary_1966}, so each is an independent set in $L_{1,n}$.
The largest VT code (corresponding to VT weight zero) always contains at least $\frac{2^n}{n+1}$ codewords.
This matches the asymptotic upper bound, so $\alpha(L_{1,n}) \sim \frac{2^n}{n}$.
The largest of these codes is conjectured to be optimal, i.e., it is conjectured to solve the maximum independent set problem on $L_{1,n}$ \cite{sloane_single-deletion-correcting_2002}. 
Kulkarni and Kiyavash~\cite{kulkarni_non-asymptotic_????} show that these codes are within a factor of at most $\frac{n+1}{n-1}$ of the largest for string length $n$.

Levenshtein also showed that the number of distinct superstrings of a string produced by $s$ insertions only depends on the length of the string \cite{levenshtein_elements_1974}.
For each $x \in [2]^{n-s}$, 
\begin{equation}
\label{eq:insertion-number}
|I_s(x)| = I_{s,n}, 
\end{equation}
where
\begin{equation*}
I_{s,n} = \sum_{i=0}^s \binom{n}{i}.
\end{equation*}
For fixed $s$, this implies
\begin{equation}
I_{s,n} \sim \binom{n}{s}.
\end{equation}

Calabi gave an upper bound on the number of substrings produced by $s$ deletions \cite{calabi_general_1969}.
For each $x \in [2]^{n+s}$, 
\begin{equation}
\label{eq:deletion-ub}
|D_s(x)| \leq I_{s,n}.
\end{equation}
For any fixed length, only the two strings of alternating zeros and ones meet this bound with equality.

\section{Code construction by weight partitioning}
\label{section:construction}
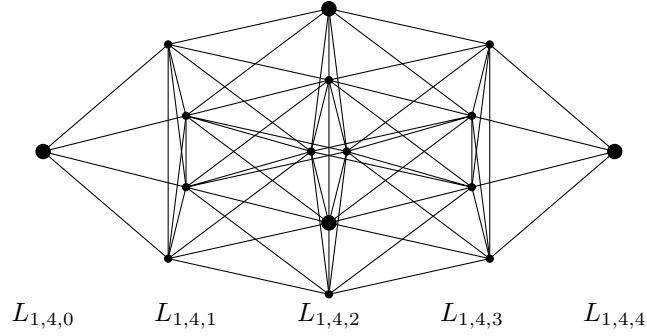
\begin{figure}
 \centering
  \begin{tikzpicture}[scale = 0.95]
 
  \coordinate (v0000) at (0.5,2);
 
  \coordinate (v0001) at (2.25,0.5);
  \coordinate (v0010) at (2.5,1.5);
  \coordinate (v0100) at (2.5,2.5);
  \coordinate (v1000) at (2.25,3.5);
  
  \coordinate (v0011) at (4.5,0);
  \coordinate (v0101) at (4.5,1);
  \coordinate (v1001) at (4.25,2);
  \coordinate (v0110) at (4.75,2);
  \coordinate (v1010) at (4.5,3);
  \coordinate (v1100) at (4.5,4);
  
  \coordinate (v0111) at (6.75,0.5);
  \coordinate (v1011) at (6.5,1.5);
  \coordinate (v1101) at (6.5,2.5);
  \coordinate (v1110) at (6.75,3.5);
 
  \coordinate (v1111) at (8.5,2);
 
  \draw (v0001) -- (v0010) -- (v0100) -- (v1000) -- (v0001);
  \draw (v1110) -- (v1101) -- (v1011) -- (v0111) -- (v1110);
  \draw (v0000) -- (v0001) -- (v0011) -- (v0111) -- (v1111);
  \draw (v0000) -- (v0010) -- (v0011) -- (v1011) -- (v1111);
  \draw (v0000) -- (v0100) -- (v1100) -- (v1101) -- (v1111);
  \draw (v0000) -- (v1000) -- (v1100) -- (v1110) -- (v1111);
  \draw (v0011) -- (v1001) -- (v1100) -- (v0110) -- (v0011);
  \draw (v0101) -- (v1001) -- (v1010) -- (v0110) -- (v0101);
  \draw (v0011) -- (v0101) -- (v1010) -- (v1100);
  \draw (v0001) -- (v0100) -- (v1010) -- (v1101) -- (v0111);
  \draw (v1000) -- (v0010) -- (v0101) -- (v1011) -- (v1110);
  \draw (v0010) -- (v1001) -- (v0100) -- (v0110) -- (v0010);
  \draw (v1101) -- (v0110) -- (v1011) -- (v1001) -- (v1101);
  \draw (v1000) -- (v1010) -- (v1110);
  \draw (v0001) -- (v0101) -- (v0111);
  \draw (v0001) -- (v1001) -- (v1000);
  \draw (v1110) -- (v0110) -- (v0111);
  \draw (v0100) -- (v0101) -- (v1101);
  \draw (v0010) -- (v1010) -- (v1011);
  
  \draw[fill] (v0000) circle [radius=0.1];
  \draw[fill] (v0001) circle [radius=0.05];
  \draw[fill] (v0010) circle [radius=0.05];
  \draw[fill] (v0100) circle [radius=0.05];
  \draw[fill] (v1000) circle [radius=0.05];
  \draw[fill] (v0011) circle [radius=0.05];
  \draw[fill] (v0101) circle [radius=0.1];
  \draw[fill] (v1001) circle [radius=0.05];
  \draw[fill] (v0110) circle [radius=0.05];
  \draw[fill] (v1010) circle [radius=0.05];
  \draw[fill] (v1100) circle [radius=0.1];
  \draw[fill] (v0111) circle [radius=0.05];
  \draw[fill] (v1011) circle [radius=0.05];
  \draw[fill] (v1101) circle [radius=0.05];
  \draw[fill] (v1110) circle [radius=0.05];
  \draw[fill] (v1111) circle [radius=0.1];
 
  \draw (0.5,-0.3) node {$L_{1,4,0}$};
  \draw (2.5,-0.3) node {$L_{1,4,1}$};
  \draw (4.5,-0.3) node {$L_{1,4,2}$};
  \draw (6.5,-0.3) node {$L_{1,4,3}$};
  \draw (8.5,-0.3) node {$L_{1,4,4}$};
  
  \end{tikzpicture}
 \caption{$L_{1,4}$ partitioned by Hamming weight. An independent set in each even weight layer is highlighted.}
 \label{figure:L14}
 \end{figure}

We now describe a strategy for code construction for any number of deletions. 
This strategy is inspired by a simple bound on deletion distance.
\begin{lemma}
\label{lemma:weight-bound}
For all strings $x,y \in [2]^n$, the deletion distance between them satisfies the lower bound $d_L(x,y)/2 \geq |H(x) - H(y)|$.
\end{lemma}
\begin{IEEEproof}
Let $z \in [2]^l$ be a longest common substring of $x$ and $y$. Then $z$ has at most as many ones than either $x$ or $y$, so
\begin{equation*}
H(z) \leq \min(H(x),H(y))
\end{equation*}
It must also have at more as many zeros, so
\begin{equation*}
l - H(z) \leq \min(n-H(x),n-H(y)).
\end{equation*}
Combining these yields
\begin{equation*}
n - l \geq \max(H(x),H(y)) - \min(H(x),H(y)).
\end{equation*}
The deletion distance is $2(n-l)$, so the claim follows.
\end{IEEEproof}
Let $L_{s,n,k}$ be the subgraph of $L_{s,n}$ induced by the vertices with exactly $k$ ones.
The endpoints of any edge in $L_{s,n}$ differ in Hamming weight by at most $s$. 
Suppose we find an independent set composed entirely of vertices of Hamming weight $k$, i.e. an independent set in $L_{s,n,k}$, 
and another independent set entirely of vertices of weight $k+s+1$, we can guarantee that their union is an independent set in $L_{s,n}$. 
Then we can add another independent set in $L_{s,n,k+2(s+1)}$ and continue until we have exhausted the weights that are equal to $k \bmod s+1$. 
This procedure gives us an independent set in $L_{s,n}$.
Fig.~\ref{figure:L14} illustrates this for $L_{1,4}$.

More formally, we have the following result. 
\begin{lemma}
\label{lemma:constant-weight-construction}
For each possible remainder $0 \leq a \leq s$, the constant weight strategy produces an $s$-deletion correcting code with at least 
$\sum_{\substack{0 \leq k \leq n   \\ k \equiv a \bmod{s+1}}} \alpha(L_{s,n,k})$ codewords.
\end{lemma}
Another way to describe this process is that we start by throwing out all the vertices whose Hamming weights do not equal $a \bmod s+1$. 
The remaining graph contains about $\frac{1}{s+1}$ of the original vertices and it is disconnected.
The maximum independent set in this graph is the union of the maximum independent sets from each connected component.

We have described how to build an independent set in $L_{s,n}$ out of independent sets in the constant weight subgraphs.
We can build a coloring of $L_{s,n}$ out of colorings of the constant weight subgraphs.
\begin{lemma}
\label{lemma:two-stage-coloring}
For $n,k \in \mathbb{N}$ with $0 \leq k \leq n$, there is some proper $c_k$-coloring of $L_{s,n,k}$, $f_k : [2]^n_k \rightarrow [c_k]$.
Then there is a coloring function
\begin{IEEEeqnarray*}{rCl}
g : [2]^n &\rightarrow& [s+1] \times [\max_k c_k]\\
x &\mapsto& (H(x) \bmod s+1, f_{H(x)}(x))
\end{IEEEeqnarray*}
that is a proper coloring of $L_{s,n}$.
\end{lemma}
\begin{IEEEproof}
Let $x$ and $y$ be adjacent vertices in $L_{s,n}$.
If $g$ is a proper coloring, it must assign them different colors.
If $H(x) = H(y)$, then $f_{H(x)}(x) \neq f_{H(x)}(y)$.
From Lemma~\ref{lemma:weight-bound}, $|H(x) - H(y)| \leq s$ so if $H(x) \neq H(y)$, then $H(x) \not\equiv H(y) \mod s+1$.
\end{IEEEproof}

\subsection{Upper Bounds on Maximum and Average Degree}
The strategy outlined above reduces the problem of finding an independent set in $L_{s,n}$ to the problem of finding independent sets in each of $L_{s,n,k}$, for $0 \leq k \leq n$.
We would like to know how the sizes of codebooks produced by the constant weight approach compare to unrestricted codes.
To make this comparison we will apply the same lower bounding technique to both types of codes.

Recall that $\alpha(G) \geq |V(G)|/(\overline{d}(G)+1)$ where $\overline{d}(G) = \frac{2|E(G)|}{|V(G)|}$, the average degree of $G$.
This translates an upper bound on average degree into a lower bound on maximum code size.
We will apply this bound to both $L_{s,n}$ and $L_{s,n,k}$.
In the case of $L_{s,n}$, we will deduce Levenshtein's original lower bound on code size.

The computation of the average degree of $L_{s,n}$ is simpler so we tackle it first.
A very similar argument applies to computing the degree of a single specified vertex so we present the two together.
\begin{lemma}
\label{lemma:degree-ubs}
For all $s,n \in \mathbb{N}$ with $s \leq n$, the average degree and maximum degree in $L_{s,n}$ satisfy
\begin{IEEEeqnarray*}{rCl}
\overline{d}(L_{s,n}) &\leq& 2^{-s} I_{s,n}(I_{s,n} - 1),\\
\overline{d}(L_{s,n}) &\lesssim& 2^{-s} \binom{n}{s}^2,\\
\Delta(L_{s,n}) &\leq& I_{s,n-s}(I_{s,n} - 1),\\ 
\Delta(L_{s,n}) &\lesssim& \binom{n}{s}^2.
\end{IEEEeqnarray*}
The asymptotic bounds are for fixed $s$.
\end{lemma}
\begin{IEEEproof}
Vertices $x$ and $y$ are adjacent if and only if $|D_s(x,y)| \geq 1$.
Thus in the whole graph we have
\begin{equation*}
|E(L_{s,n})| = \sum_{x,y \in [2]^n, x \neq y} \min(|D_s(x,y)|,1).
\end{equation*}
We can count the triples $x,y \in [2]^n$, $z~\in~[2]^{n-s}$ such that $x > z$ and $y > z$ in two ways.
On the left we sum over $x$ and $y$ and on the right we sum over $z$:
\begin{IEEEeqnarray*}{rCl}
\sum_{x,y \in [2]^n, x \neq y} |D_s(x,y)| &=& \sum_{z \in [2]^{n-s}} \binom{|I_s(z)|}{2}, \\
&=& 2^{n-s} \binom{I_{s,n}}{2}.
\end{IEEEeqnarray*}
Recall that $|I_s(z)|$ is a constant equal to $I_{s,n}$ from \eqref{eq:insertion-number}.
The average degree is given by $\overline{d}(L_{s,n}) = \frac{2 |E(L_{n,s})|}{|V(L_{n,s})|}$, so
\begin{equation*}
\overline{d}(L_{s,n}) \leq \frac{2^{n-s+1}}{2^n} \binom{I_{s,n}}{2} = 2^{-s} I_{s,n}(I_{s,n}-1) \sim 2^{-s} \binom{n}{s}^2.
\end{equation*}

To prove the bounds on maximum degree, we consider the neighborhood of a vertex instead of the entire graph.
We have
\begin{equation*}
|N_s(x)| = \sum_{y \in [2]^n \setminus x} \min(|D_s(x,y)|,1)
\end{equation*}
and
\begin{IEEEeqnarray*}{rCl}
\sum_{y \in [2]^n \setminus x} |D_s(x,y)| &=& \sum_{z \in D_s(x)}|I_s(z) \setminus x|,\\
&=& |D_s(x)|(I_{s,n} - 1),\\
&\leq& I_{s,n-s}(I_{s,n}-1) .
\end{IEEEeqnarray*}
The inequality follows from \eqref{eq:deletion-ub} in Section~\ref{subsection:existing}.
Thus the maximum degree satisfies
\begin{equation*}
\Delta(L_{s,n}) \leq I_{s,n-s}(I_{s,n}-1) \sim \binom{n}{s}^2.
\end{equation*}

\end{IEEEproof}
Levenshtein's original lower bound follows immediately.
\begin{theorem}
For all $s,n \in \mathbb{N}$, there exist codebooks of size at least $\frac{2^{n+s}}{I_{s,n}(I_{s,n} - 1) + 2^s}$.
For fixed $s$, their size is asymptotically at least $\frac{2^{n+s}}{\binom{n}{s}^2}$.
\end{theorem}
\begin{IEEEproof}
Codes are independent sets in $L_{s,n}$.
Substituing the upper bound on $\overline{d}(L_{s,n})$ of Lemma~\ref{lemma:degree-ubs} into Turan's theorem, $\alpha(G) \geq |V(G)|/(\overline{d}(G)+1)$, gives the result.
\end{IEEEproof}
Levenshtein's original proof of the asymptotic version of this result used a different argument \cite{levenshtein_binary_1966}.
He later proved the nonasymptotic version using what appears to be the same argument that we make here \cite{levenshtein_bounds_2002}.

\subsection{Lower Bounds on Sizes of Code from the Constant Weight Strategy}
\label{subsection:lower-bounds}

Now we extend this argument to the constant weight strategy.
We used the total number of superstrings of a string to bound the average degree of $L_{s,n}$,
and we will use the number of superstrings of a given weight to bound the average degree of $L_{s,n,k}$.
This will translate into a bound on the size of independent sets in $L_{s,n,k}$ and independent sets in $L_{s,n}$ with our weight restriction.
We need some additional notation.
\begin{mydef}
For $x \in [2]^n_k$, let
\begin{equation*}
I_{(s,r)}(x) = \left\{ \left. w \in [2]^{n+s}_{k+r} \right| w > x \right\}.
\end{equation*}
This is the set of superstrings of $x$ with length $n+s$ and weight $k+r$, the superstrings produced by inserting $r$ ones and $s-r$ zeros.
\end{mydef}

Just as the size of $I_s(x)$ only depends on the length of $x$, the size of $I_{(s,r)}(x)$ only depends on the length and weight of $x$.
\begin{lemma}
\label{lemma:layer-insertion-number}
For all $n,k,s,r \in \mathbb{N}$ with $0 \leq r \leq s \leq n$ and $0 \leq k \leq n$, and all $x \in [2]^{n-s}_{k-r}$, the number of superstrings of $x$ with length $n$ and weight $k$ satisfies $|I_{(s,r)}(x)| =  \sum_{i=0}^{\min(r,s-r)} \binom{k+s-2r}{s-r-i} \binom{n-k-s+2r}{r-i}$.
\end{lemma}
The proof is quite involved; it requires a new representation of the elements of $I_{(s,r)}(\cdot)$ in terms of multisets. So as to not hinder the flow of our results, we have included it in Appendix~\ref{section:counting-superstrings}.

We will name this constant:
\begin{equation}
I_{(s,r),(n,k)} = \sum_{i=0}^{\min(r,s-r)} \binom{k+s-2r}{s-r-i} \binom{n-k-s+2r}{r-i}
\end{equation}
For the following lemma, we need the asymptotic value of this expression letting $k = pn$ with fixed $s$, $r$, and $p$.
The $i=0$ term of the sum is a degree $s$ polynomial and all other terms are of lower degree. 
Thus we have
\begin{IEEEeqnarray}{rCl}
I_{(s,r),(n,pn)} &\sim& \binom{pn+s-2r}{s-r} \binom{n-pn-s+2r}{r}, \nonumber \\
&\sim& \binom{pn}{s-r} \binom{n-pn}{r}, \nonumber \\
&\sim& \frac{(pn)^{s-r}}{(s-r)!} \frac{(n-pn)^r}{r!}, \nonumber \\
\label{eq:weight-insertion-number}
&\sim&\binom{n}{s}\binom{s}{r}p^{s-r}(1-p)^r.
\end{IEEEeqnarray}

We will use the following lemma in our computation of the average degree of the constant weight subgraphs.
\begin{lemma}
\label{lemma:sum-bound}
For all $s \in \mathbb{N}$,
\begin{equation*}
f_s(p) = \sum_{0 \leq r \leq s} \binom{s}{r}^2 p^{s-r}(1-p)^r,
\end{equation*}
is maximized at $p = 1/2$, so for all $p$, $f_s(p) \leq 2^{-s} \binom{2s}{s}$.
\end{lemma}
The proof is in Appendix~\ref{section:proof-of-sum-bound}.
The following lemma gives the average degree of $L_{s,n,k}$.

\begin{lemma}
\label{lemma:layer-average-degree}
Let $k = pn$. Then the average degree of the weight $k$ subgraph satisfies $\overline{d}(L_{s,n,k}) \lesssim \left( \frac{p(1-p)}{2} \right)^s\binom{2s}{s} \binom{n}{s}^2 $.
\end{lemma}
\begin{IEEEproof}
Let $x$ be a string of length $n-s$ and weight $k-r$ for some $0 \leq r \leq s$.
Any two vertices in $I_{(s,r)}(x)$ are adjacent in $L_{s,n,k}$.
There are $\binom{n-s}{k-r} \binom{I_{(s,r),(n,k)}}{2}$ such pairs of vertices.
The endpoints of each edge in $L_{s,n,k}$ have at least one common substring of length $n-s$.
The weight of this substring must be $k-r$ for some $0 \leq r \leq s$ because at most $s$ ones were deleted from $x$ or $y$ to produce it.
Thus every edge is counted at least once in the sum in
\begin{equation}
\label{eq:ub1}
\frac{2 |E(L_{n,s,k})|}{|V(L_{n,s,k})|} \leq \frac{2}{\binom{n}{k}} \sum_{0 \leq r \leq s} \binom{n-s}{k-r} \binom{I_{(s,r),(n,k)}}{2}.
\end{equation}
Recall that for fixed $a$, $\binom{x}{a} \sim \frac{x^a}{a!}$.
The ratio of binomial coefficients simplifies asymptotically to
\begin{equation}
\label{eq:binom-ratio}
\frac{\binom{n-s}{k-r}}{\binom{n}{k}} = \frac{\binom{k}{r}\binom{n-k}{s-r}}{\binom{n}{s}\binom{s}{r}} \sim \frac{k^r(n-k)^{s-r}}{n^s} = p^r(1-p)^{s-r} .
\end{equation}
Substituting \eqref{eq:weight-insertion-number} and \eqref{eq:binom-ratio} into \eqref{eq:ub1} gives an asymptotic upper bound on $\overline{d}(L_{s,n,k})$ of
\begin{IEEEeqnarray*}{rCl}
\overline{d}(L_{s,n,k}) & \lesssim & \sum_{0 \leq r \leq s} p^r(1-p)^{s-r} \left( \binom{n}{s} \binom{s}{r} p^{s-r} (1-p)^r \right)^2, \\
&=& p^s(1-p)^s \binom{n}{s}^2 \sum_{0 \leq r \leq s} \binom{s}{r}^2 p^{s-r}(1-p)^r,\\
\end{IEEEeqnarray*}
Applying Lemma~\ref{lemma:sum-bound} give the final bound.
\begin{equation*}
\overline{d}(L_{s,n,k}) \lesssim p^s(1-p)^s \binom{n}{s}^2 2^{-s} \binom{2s}{s}.
\end{equation*}
\end{IEEEproof}
We can now use the upper bound on average degree to get a lower bound on code size.
\begin{theorem}
\label{theorem:constant-weight-code-sizes}
For fixed $s$, $s$-deletion correcting codes produced by the constant weight strategy contain asymptotically at least $\frac{2^{n+3s}}{(s+1)\binom{2s}{s}\binom{n}{s}^2}$ codewords.
\end{theorem}
\begin{IEEEproof}
From Lemma~\ref{lemma:constant-weight-construction} there is a code with at least 
\begin{equation*}
\sum_{\substack{0 \leq k \leq n   \\ k \equiv a \bmod{s+1}}} \alpha(L_{s,n,k}) \geq
\frac{1}{s+1}\sum_{k = 0}^n \alpha(L_{s,n,k})
\end{equation*}
 codewords.
The inequality holds because for some $a$ the resulting code is at least as large as the average.
By Turan's Theorem, $\alpha(L_{s,n,k}) \geq |V(L_{s,n,k})|/(\overline{d}(L_{s,n,k}) + 1)$.
Taking the result of Lemma~\ref{lemma:layer-average-degree} and applying $p(1-p) \leq 1/4$ gives $\overline{d}(L_{s,n,k}) + 1 \lesssim \binom{2s}{s} \binom{n}{s}^2 /2^{3s}$.
This bound does does not depend on $k$, so using $\sum_{k=0}^n |V(L_{s,n,k})| = 2^n$ completes the proof.
\end{IEEEproof}

\begin{corollary}
The size of codebooks produced by the constant weight strategy is a factor of $\frac{(s+1) \binom{2s}{s}}{2^{2s}} \leq \frac{s+1}{\sqrt{2s}}$ below the Levenshtein lower bound.
\end{corollary}
\begin{IEEEproof}
The ratio is 
\begin{equation*}
\left. \frac{2^{n+s}}{\binom{n}{s}^2} \right/ \frac{2^{n+3s}}{(s+1)\binom{2s}{s}\binom{n}{s}^2} = \frac{(s+1) \binom{2s}{s}}{2^{2s}}.
\end{equation*}
From Stirling's approximation, $\binom{2s}{s} \leq \frac{2^{2s}}{\sqrt{2s}}$.
The result is immediate.
\end{IEEEproof}

\subsection{Algorithms}
In this section we will compare the algorithms that produce optimal codes, codes promised by Turan's theorem, and explicit codes.
Computing the size of the largest independent set is NP-hard for general graphs.
The best known exact algorithm that requires only polynomial space uses $O(poly(n)2^{0.288n})$ time, where $n$ is the number of vertices. \cite{fomin_measure_2009}.
\begin{theorem}
The ratio of the upper bound on the run time of the best exact algorithm on $L_{s,n}$ to the sum of upper bounds for the run time on each of the graphs $L_{s,n,k}$ is $\Theta(poly(n)2^{0.288(1-\sqrt{2/\pi n})2^n})$.
\end{theorem}
\begin{IEEEproof}
For each $n$ and $s$, there are only $n/2$ different graphs $L_{s,n,k}$, so running the algorithm on all of them takes at most a linear factor longer than running the algorithm on the largest of them.
The largest constant weight graph, $L_{s,n,n/2}$, contains $\binom{n}{n/2}$ vertices.
By Stirling's approximation this is asymptotically $\sqrt{\frac{2}{\pi n}}2^n$.
Thus the total run time is at most $O(poly(n)2^{0.288\sqrt{2/\pi n}(2^n)})$.
The run time for $L_{s,n}$ is at most $O(poly(n)2^{0.288(2^n)})$.
\end{IEEEproof}

However, the number of vertices in $L_{s,n,n/2}$ is still exponential in $n$ so exact algorithms quickly become infeasible.

There are many classes of graphs for which faster algorithms exist, but we have not found such a class that contains $\{L_{s,n}|s,n \in \mathbb{N}\}$.
One of the most general such classes is the class of perfect graphs.
\begin{theorem}
\label{theorem:imperfect}
For all $s,n \in \mathbb{N}$ with $s \geq 1$ and $n \geq 3s+1$, $L_{s,n}$ is not a perfect graph.
\end{theorem}
The proof is in Appendix~\ref{section:induced}; it involves showing that there are odd cycles with no chords in $L_{s,n}$.

The independent sets promised by Turan's Theorem (i.e. by Theorem~\ref{theorem:constant-weight-code-sizes}) can be found by a greedy algorithm using a minimum degree heuristic \cite{west_introduction_2001}.
Greedy codes can be generated in time polynomial in the the number of vertices in the graph.
Every vertex in $L_{s,n}$ is in some $L_{s,n,k}$, so there is no time advantage to running a greedy algorithm on all of $L_{s,n,k}$ over running it on $L_{s,n}$. 

The number of vertices in $L_{s,n}$ is exponential in $n$, so even the greedy algorithms are slow.
Because the independent sets that we seek contain exponentially many vertices, listing the members of a set is slow regardless of the complexity of the algorithm that we use to find the set.
This difficulty leads to our interest in explicit codes, which satisfy an even stronger algorithmic condition.
To demonstrate the difference between a greedy code and an explicitly constructed code, consider an independent set $S$ in $G$ as the indicator function $\mathbf{1}_S:V(G) \rightarrow [2]$.
In an explicit code, one can compute this function to test membership code quickly and in small space.
In contrast, to test membership in a greedy code one can store the set of codewords and search, which requires space exponential in $n$, or regenerate the code, which requires time exponential in $n$.

A $k$-coloring of a graph $G$ is naturally thought of as a function $f:V(G) \rightarrow [k]$.
An easy to compute coloring function leads immediately to an easy to compute indicator function.
In the following section we show an explicit construction of a single deletion correcting code using the constant weight approach.
The weight condition together with a simple coloring function allow membership testing of a vertex in time and space linear in $n$.

\section{Single deletion construction}
\label{section:single-deletion}
In this section, we focus on the single deletion case ($s=1$).
We show an explicit construction of independent sets in the graphs $L_{1,n,k}$.
We construct these independent sets by finding a coloring of $L_{1,n,k}$.
This coloring is closely related to the VT codes in $L_{1,n}$.
The code that results from our coloring is asymptotically optimal.

\subsection{An explicit coloring of the constant weight single deletion graphs}
The VT construction uses a weight function to partition $[2]^n$ into $n+1$ codes, each an independent set in $L_{1,n}$.
We observe that this makes the VT weight a proper $(n+1)$-coloring of $L_{1,n}$, so $\chi(L_{1,n}) \leq n+1$.

Both the VT coloring of $L_{1,n}$ and our colorings of $L_{1,n,k}$ are based on the following weight function.
\begin{mydef}
For any $x \in [2]^n$, let $w(x) = \sum_{i=0}^{n-1} (i~+~1)x_i$.
Call $w(x) \bmod n+1$ the VT weight.
Let $f_k(x) = w(x) \bmod{(\max(k,n-k)+1)}$.
We call $f_k$ the modified VT weight.
\end{mydef}

Levenshtein showed that for each string length $n$, the Varshamov-Tenengolts construction provides $n+1$ distinct single deletion correcting codes \cite{levenshtein_binary_1966}. 
Restated in terms of graphs, the VT weight is a proper coloring of $L_{1,n}$.

\begin{lemma}
\label{lemma:layer-coloring}
The modified VT weight $f_k$ is a proper coloring of $L_{1,n,k}$.
\end{lemma}
\begin{IEEEproof}
Let $x$ and $y$ be adjacent vertices in $L_{1,n,k}$.
We will show that $f_k(x) \neq f_k(y)$.
Index the symbols in $x$ and $y$ by $[n]$, so $x=(x_0,..,x_{n-1})$.
For $S \subset [n]$, let $x_S$ indicate the substring of $x$ consisting of the symbols whose indices are in $S$.

Note that $\sum_{i=0}^{n-1} x_i = \sum_{i=0}^{n-1} x_i = k$, so 
\begin{equation*}
w(y) - w(x) = \sum_{i=0}^{n-1} (i+1)(y_i - x_i) = \sum_{i=0}^{n-1} i(y_i - x_i).
\end{equation*}
Let $a$ be the smallest index where $x_a \neq y_a$ and let $b$ be the largest such index, so
\begin{equation*}
w(y) - w(x) = \sum_{i=a}^{b} i(y_i - x_i).
\end{equation*}
Because $d_L(x,y) = 1$, $x$ and $y$ have a common substring $z$ of length $n-1$.
Either $z = x_{[n] \setminus a} = y_{[n] \setminus b}$ or $z = x_{[n] \setminus b} = y_{[n] \setminus a}$.
Without loss of generality assume the latter.
Then for $a \leq i \leq b-1$, $z_i = x_i = y_{i+1}$.
Because $H(x) = H(y) = k$, we have $x_b = y_a$.
\begin{IEEEeqnarray*}{rCl}
w(y) - w(x) & = & ay_a - bx_b + \sum_{i=a}^{b-1} (i+1) y_{i+1} - ix_i \\
& = & (a-b)x_b + \sum_{i=a}^{b-1} x_i \\
\end{IEEEeqnarray*}
Let $l = \sum_{i=a}^{b-1} x_i$, the number of ones in $x_{\{i..j-1\}}$.
There are two cases to consider, $x_b = 0$ and $x_b = 1$.
If $x_b = 0$, then $w(y) - w(x) = l$. 
Since $x \neq y$, $x_a = 1$ and $0 < l \leq k$. 
If $x_b = 1$, then $w(x) - w(y) = b-a-l$, the number of zeros in $x_{\{a..b-1\}}$.
Since $x \neq y$, $x_a = 0$ and $0 < b-a-l \leq n-k$.
In both cases, $0 < |w(y) - w(x)| \leq \max(k,n-k)$, so $w(x) \bmod{(\max(k,n-k)+1)} \neq 0$.
\end{IEEEproof}

\subsection{Lower bounds on coloring single deletion graphs}
\label{subsection:coloring-single}
We show that both the VT coloring of $L_{1,n}$ and our coloring of $L_{1,n,k}$ are optimal.
In both cases, to demonstrate optimality, we will find cliques of matching size.
Recall that the vertices in a clique must each be assigned different colors, so $\omega(G) \leq \chi(G)$.
The following lemma constructs these cliques.

\begin{lemma}
\label{lemma:cliques}
For all $s,n \in \mathbb{N}$, $s \leq n$, $L_{s,n}$ contains cliques of $I_{s,n}$ vertices.
For all $r,s,k,n \in \mathbb{N}$ such that $0 \leq r \leq s$ and $r \leq k \leq n - s + r$, $L_{s,n,k}$ contains cliques of $I_{(s,r),(n,k)}$ vertices.
\end{lemma}
\begin{IEEEproof}
By \eqref{eq:insertion-number}, each string in $[2]^{n-s}$ has $I_{s,n}$ superstrings in $[2]^n$.
These are all adjacent in $L_{s,n}$, so they form a clique.
By Lemma~\ref{lemma:layer-insertion-number}, each string in $[2]^{n-s}_{k-r}$ has $I_{(s,r),(n,k)}$ superstrings in $[2]^n_k$. 
These are all adjacent in $L_{s,n,k}$, so they form a clique.
\end{IEEEproof}
The optimality of both colorings follows immediately.
\begin{theorem}
For all $n$, the VT coloring of $L_{1,n}$ is optimal and 
\begin{equation*}
\chi(L_{1,n}) = \omega(L_{1,n}) = n+1.
\end{equation*}
For all $n$ and $1 \leq k \leq n-1$, the coloring of $L_{1,n,k}$ by the modified VT weight $f_k$ is optimal and 
\begin{equation*}
\chi(L_{1,n,k}) = \omega(L_{1,n,k}) = \max(k,n-k)+1.
\end{equation*}
\end{theorem}
\begin{IEEEproof}
By Lemma~\ref{lemma:cliques}, $L_{1,n}$ contains cliques of $I_{1,n} = n+1$ vertices.
The VT coloring uses $n+1$ colors, so $n+1 \leq \omega(L_{1,n}) \leq \chi(L_{1,n}) \leq n+1$.

By Lemma~\ref{lemma:cliques}, $L_{1,n,k}$ contains cliques of sizes $I_{(1,0),(n,k)} = k+1$ and $I_{(1,1),(n,k)} = n-k+1$. 
From Lemma~\ref{lemma:layer-coloring} we have $\max(k,n-k)+1 \leq \omega(L_{1,n,k}) \leq \chi(L_{1,n,k}) \leq \max(k,n-k)+1$.
\end{IEEEproof}

\subsection{Asymptotic optimality of our codes}
We now show that taking the union of independent sets from $L_{1,n,k}$ produces an independent set in $L_{1,n}$ that is asymptotically of optimal size. 
Let $C_{n,k}$ be a largest color class of $L_{1,n,k}$ using the coloring described above. 
For $a \in [2]$, our code is the set $D_{n,a}$,
\[
D_{n,a} := \bigcup_{\substack{0 \leq k \leq n   \\ k \equiv a \bmod{2}}} C_{n,k}.
\]

\begin{lemma}
$|D_{n,a}| \geq \frac{1}{n+1} \left( 2^n - \binom{n}{k^*} \right)$ where $k^*$ is the integer closest to $n/2$ such that $k^* \not\equiv a \bmod{2}$.
\end{lemma}
\begin{IEEEproof}
In each graph $L_{1,n,k}$, some color class must be at least as large as the average, so
\begin{equation*}
|D_{n,a}| = 
\sum_{\substack{0 \leq k \leq n \\ k \equiv a \bmod{2}}} |C_{n,k}| 
\geq \sum_{\substack{0 \leq k \leq n   \\ k \equiv a \bmod{2}}} \frac{|V(L_{1,n,k})|}{\chi(L_{1,n,k})}.
\end{equation*}
There are $\binom{n}{k}$ vertices in $L_{1,n,k}$ and from Lemma~\ref{lemma:layer-coloring} we have $\chi(L_{1,n,k}) \leq \max(k,n-k)+1$. 
Thus $|D_{n,a}|$ is at least
\begin{IEEEeqnarray*}{Cl}
& \sum_{\substack{k = 0 \\ k \equiv a \bmod{2}}}^{k^*-1} \binom{n}{k}\frac{1}{n-k+1} 
     + \sum_{\substack{k = k^*+1 \\ k \equiv a \bmod{2}}}^n \binom{n}{k}\frac{1}{k+1}, \\
= & \sum_{\substack{k = 0 \\ k \equiv a \bmod{2}}}^{k^*-1} \binom{n+1}{k}\frac{1}{n+1} 
     + \sum_{\substack{k = k^*+1 \\ k \equiv a \bmod{2}}}^n \binom{n+1}{k+1}\frac{1}{n+1}. \\
\end{IEEEeqnarray*}
Because $\binom{n+1}{k} = \binom{n}{k-1} + \binom{n}{k}$, we can rewrite the lower bound as 
\begin{IEEEeqnarray*}{Cl}
 & \frac{1}{n+1}\left(\sum_{k = 0}^{k^*-1} \binom{n}{k}
     + \sum_{k = k^*+1}^n \binom{n}{k}\right), \\
= & \frac{1}{n+1}\left(2^n-\binom{n}{k^*}\right).
\end{IEEEeqnarray*}
\end{IEEEproof}

\begin{theorem}
The sequence of codes $D_{n,a}$ is asymptotically optimal.
\end{theorem}
\begin{IEEEproof}
By Stirling's formula, $\binom{n}{n/2} \sim 2^n\sqrt{\frac{2}{\pi n}}$, so 
\begin{equation*}
|D_{n,a}| \sim \frac{2^n}{n+1} \left( 1-\sqrt{\frac{2}{\pi n}} \right) \sim \frac{2^n}{n}.
\end{equation*}
Recall from Section~\ref{subsection:existing} that $\alpha(L_{1,n}) \sim \frac{2^n}{n}$, so the code is asymptotically optimal.
\end{IEEEproof}

Note that $\max_k \chi(L_{1,n,k}) = n$, so the colorings constructed by Lemma~\ref{lemma:two-stage-coloring} use $2n$ colors.
They are far from optimal because $\chi(L_{1,n})=n+1$. 
However, most of the vertices are in the subgraphs with $k \approx n/2$, and $\chi(L_{1,n,n/2}) = n/2 + 1$. 
Thus, half the vertices have been thrown out, but the middle layers are colored about twice as efficiently as they were in the original graph.
There are $2n$ color classes, but their size vary significantly and only $n+2$ of them contain the most of the vertices.
This explains the asymptotic optimality.

\section{Lower bounds on coloring $L_{s,n}$}
\label{section:coloring}
In this section, we will show that for correcting multiple deletions, the independent sets guaranteed by an asymptotically optimal coloring do not match the Levenshtein upper bound.
This means that either solving the coloring problem does not guarantee a solution to the independent set problem or the Levenshtein upper bound is not tight.

More concretely, we show that $\chi(L_{s,n}) \gtrsim \binom{n}{s}\binom{s}{\lfloor s/2 \rfloor}$ whereas for the average size of the color classes to match Levenshtein's upper bound, we need $\chi(L_{s,n}) \sim \binom{n}{s}$.

In Section~\ref{section:preliminaries} we gave two lower bounds on chromatic number for any graph $G$.
First, $\chi(G) \geq |V(G)|/\alpha(G)$.
Levenshtein's asymptotic upper bound is $\alpha(L_{s,n}) \lesssim 2^n/\binom{n}{s}$ \cite{levenshtein_binary_1966}.
Combining these yields $\chi(L_{s,n}) \gtrsim \binom{n}{s}$.
Second, $\chi(G) \geq \omega(G)$.
From Lemma~\ref{lemma:cliques}, we know that cliques in $L_{s,n}$ produced by a single common substring contain $I_{s,n}$ vertices and $I_{s,n} \sim \binom{n}{s}$.
Again we get $\chi(L_{s,n}) \gtrsim \binom{n}{s}$.

In general, if the first bound is tight ($\alpha(L_{s,n}) \chi(L_{s,n}) \sim |V(L_{s,n})|$), then solving the coloring problem leads to many asymptotically optimal codes.
For any asymptotically optimal sequence of colorings, almost all sequences of color classes are asymptotically optimal sequences of independent sets.
In the single deletion case, this is the case and consequently one might hope that the same is true for all $s$.
However, for multiple deletions we improve the second lower bound on chromatic number by showing that $\omega(L_{s,n}) \gtrsim \binom{n}{s}\binom{s}{\lfloor s/2 \rfloor}$.

Consequently, average sized color classes in an optimal coloring of $L_{s,n}$ do not meet Levenshtein's upper bound on $\alpha(L_{s,n})$.
If we only know the chromatic number of $L_{s,n}$, we can only guarantee the existance of color classes of the average size, $2^n/\chi(L_{s,n})$ vertices.
It is possible that there are optimal colorings in which the size of the largest color class is much larger than the average size.
It is also possible that average sized color classes in an optimal coloring are asymptotically optimal independent sets because the Levenshtein upper bound is not tight.

\subsection{Large cliques and high degree vertices}
To improve the lower bound on the chromatic number of $L_{s,n}$, we need to find large cliques in $L_{s,n}$.
That is the goal of this section.

In $L_{1,n}$, cliques produces by a single common substring are maximum, but in $L_{s,n}$ for $s \geq 2$ this a more general construction produces larger cliques.
For any string $x$ of length $m$, consider all of the strings of length $n$ within deletion a deletion distance of $s$.
By the triangle inequality, the deletion distance between any two of these strings is at most $2s$, so they form a clique in $L_{s,n}$.
If we let $m = n-s$, then every string in the clique has $x$ as a substring.
If we let $m = n+s$, then every string in the clique has $x$ as a superstring.
Bigger cliques can be constructed by letting $m$ be closer to $n$.
Recall from Section~\ref{subsection:graph} that $\omega(G^2) \geq \Delta(G)+1$ for any graph $G$ because the neighborhood of any vertex in $G$ is a clique in $G^2$.
When $s$ is even, we can let $m = n$.
In this case, $x$ is also a vertex in $L_{s,n}$ and we are effectively applying the bound.

\begin{lemma}
\label{lemma:concat-substrings}
For any strings $x,x',y,y'$, not necessarily of the same length, $d_L(x x', y y') \leq d_L(x,y) + d_L(x',y')$.
\end{lemma}
\begin{IEEEproof}
Let $|x|$ denote the length of $x$.
The strings $x$ and $y$ have a common substring $z$ of length $(|x| + |y| - d_L(x,y))/2$ and 
$x'$ and $y'$ have a common substring $z'$ of length $(|x'| + |y'| - d_L(x',y'))/2$.
The string $zz'$ is a common substring of $xx'$ and $yy'$, so the claimed bound holds.
\end{IEEEproof}

\begin{figure}
\centering
 \begin{tikzpicture}[
 scale = 1/3,
 b/.style={very thick}]

 \draw    (0, 2) grid      (24,3);
 \draw[b] (0, 2) rectangle (6, 3);
 \draw[b] (6, 2) rectangle (9, 3);
 \draw[b] (9, 2) rectangle (15,3);
 \draw[b] (15,2) rectangle (18,3);
 \draw[b] (18,2) rectangle (24,3);

 \draw    (0, 0) grid      (24,1);
 \draw[b] (0, 0) rectangle (5, 1);
 \draw[b] (5, 0) rectangle (8, 1);
 \draw[b] (8, 0) rectangle (14,1);
 \draw[b] (14,0) rectangle (17,1);
 \draw[b] (17,0) rectangle (24,1);

 \draw[b] (0, 1) -- (0, 2);
 \draw[b] (5, 1) -- (6, 2);
 \draw[b] (8, 1) -- (9, 2);
 \draw[b] (14,1) -- (15,2);
 \draw[b] (17,1) -- (18,2);
 \draw[b] (24,1) -- (24,2);

 \foreach \i in {0,2,4,6,7,8,10,12,14,18,20,22}
    \draw (\i + 0.5, 2.5) node {0};
 \foreach \i in {1,3,5,9,11,13,15,16,17,19,21,23}
    \draw (\i + 0.5, 2.5) node {1};

 \foreach \i in {0,2,3,5,6,7,9,11,13,17,19,20,22}
    \draw (\i + 0.5, 0.5) node {0};
 \foreach \i in {1,4,8,10,12,14,15,16,18,21,23}
    \draw (\i + 0.5, 0.5) node {1};

 \draw (-1,2.5) node {$x:$};
 \draw (-1,0.5) node {$y:$};
 \draw (3,3.5) node {A};
 \draw (12,3.5) node {A};
 \draw (21,3.5) node {A};
 \draw (2.5,-0.5) node {C};
 \draw (11,-0.5) node {A};
 \draw (20.5,-0.5) node {B};

 \end{tikzpicture}

\caption{
Two of the strings constructed in the proof of Lemma~\ref{lemma:clique-vertex-construction}:
the center string $x$, which has segments of types AAA, and a string $y \in S$, which has segments of types CAB.
The parameters are $l=6$, $k=3$, and $b = c = 1$.
}
\label{figure:strings}
\end{figure}
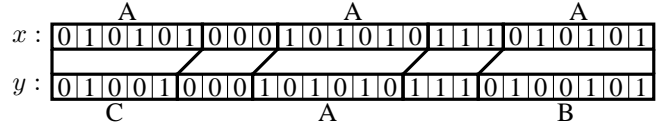

\begin{lemma}
\label{lemma:clique-vertex-construction}
For all $n \in \mathbb{N}$, the maximum clique size in $L_{s,n}$ satisfies $\omega(L_{s,n}) \gtrsim \binom{n}{s}\binom{s}{\lfloor s/2 \rfloor}$
and the maximum degree in $L_{s,n}$ satisfies $\Delta(L_{s,n}) \sim \binom{n}{s}^2$.
\end{lemma}
\begin{IEEEproof}
For all $b,c,k,l \in \mathbb{N}$ with $b + c \leq k$, let $m = k(l + 3) - 3$ and $n = m + b - c$.
We will construct a string $x \in [2]^m$ and a set $S \subset [2]^m$ such that $|S| = \binom{k}{s}\binom{s}{r}l^b(l-2)^c$ and for all $y \in S$, $d_L(x,y) \leq b + c$.
We will specify each of these strings by their pattern of runs.
All of these strings have the same first bit.
All contain $k$ segments separated by runs of length 3, so the separator differs from the last bit of the previous segment as well as the first bit of the next segment.

We will use three types of segments, types A, B and C.
Segments of type A have length $l$ and consist of runs of length 1.
Segments of type B have total length $l+1$ and contain one run of length 2 and $l-1$ runs of length 1.
There are $l$ possible run patterns with this distribution.
Each segment of type B is a superstring of the segment of type A.
Segments of type C have total length $l-1$ and contain one run of length 2 and $l-3$ runs of length 1.
There are $l-2$ possible run patterns with this distribution.
Each segment of type B is a substring of the segment of type A.

In $x$, all $k$ segments are of type A.
In an element of $S$, there are $k-a-b$ segments of type A, $b$ of type B, and $c$ of type C.
Thus there are $\binom{k}{b,c,k-b-c}$ possible sequences of the types and $\binom{k}{b,c,k-b-c}l^b(l-2)^c$ elements of $S$.
Fig.~\ref{figure:strings} gives an example.

Now we need to show that for all $y \in S$, $d_L(x,y) \leq b+c$.
The number of runs within a segment is always $l$ or $l-2$, so the boundary runs of length 3 have the same compositions in $x$ and all elements of $S$.
In any $y \in S$, there are $b+c$ segments that differ from $x$.
In each case, the deletion distance between the segment in $x$ and the segment in $y$ is one.
The rest of the strings match exactly.
By Lemma~\ref{lemma:concat-substrings}, $d_L(x,y) \leq b+c$.

Taking $k \sim l \sim \sqrt{n}$ yields 
\begin{equation*}
|S| \sim \frac{k^{b+c}}{b!c!}l^{b+c} \sim \frac{n^{b+c}}{b!c!} \sim \binom{n}{b}\binom{n}{c}.
\end{equation*}

By the triangle inequality, for all $y,z \in S$, $d_L(y,z) \leq 2(b+c)$.
Thus the vertices in $S$ form a clique in $L_{b+c,n}$.
To maximize the size of this clique for a given $s$, let $b = \lfloor s/2 \rfloor$ and $c = \lceil s/2 \rceil$.
Then
\begin{equation*}
\omega(L_{s,n}) \gtrsim \binom{n}{\lfloor s/2 \rfloor}\binom{n}{\lceil s/2 \rceil} \sim \binom{n}{s}\binom{s}{\lfloor s/2 \rfloor}.
\end{equation*}

If we let $b = c = s$, then $x$ and all elements of $S$ are the same length.
Thus $x$ is a vertex in $L_{s,n}$ and $S$ is a subset of its neighborhood.
The degree of $x$ in $L_{s,n}$ is at least $|S|$, so 
\begin{equation*}
\Delta(L_{s,n}) \gtrsim \binom{n}{s}^2.
\end{equation*}
From Lemma~\ref{lemma:degree-ubs} we have $\Delta(L_{s,n}) \lesssim \binom{n}{s}^2$.
\end{IEEEproof}

\begin{corollary}
\label{corollary:chi}
For all $s,n \in \mathbb{N}$ with $s \leq n$, the chromatic number of $L_{s,n}$ satisfies $\chi(L_{s,n}) \gtrsim \binom{n}{s}\binom{s}{\lfloor s/2 \rfloor}$.
\end{corollary}
\begin{IEEEproof}
This follows from the basic inequality $\chi(L_{s,n}) \geq \omega(L_{s,n})$.
\end{IEEEproof}
This leads us to the main theorem of this section.
\begin{theorem}
For all fixed $s \in \mathbb{N}$ with $s \geq 2$, the following inequalities hold but at most one is tight:
\begin{equation*}
\frac{2^n}{\chi(L_{s,n})} \lesssim \alpha(L_{s,n}) \lesssim \frac{2^n}{\binom{n}{s}}.
\end{equation*}
\end{theorem}
\begin{IEEEproof}
The lower bound follows from $\alpha(G)\chi(G) \geq |V(G)|$.
The upper bound is Levenshtein's.
From Corollary~\ref{corollary:chi}, 
\begin{equation*}
\frac{2^n}{\chi(L_{2s,n})} \lesssim \frac{2^n}{\binom{n}{s}\binom{s}{\lfloor s/2 \rfloor}}. 
\end{equation*}
For $s \geq 2$, $\binom{s}{\lfloor s/2 \rfloor} \geq 2$.
\end{IEEEproof}
Thus knowing the asymptotic behavior of $\chi(L_{s,n})$ does not give the asymptotic behavior of $\alpha(L_{s,n})$.

\section{Conclusion}
We investigated two approaches to code construction.
We showed that a two stage approach that restricts the weight of codewords trades a small penalty in guaranteed code size for a large reduction in computational complexity of construction.
This approach produces a new single deletion correcting code that is asymptotically optimal.

The second approach that we investigated is code construction via graph coloring.
The VT codes are an optimal coloring of the whole single deletion graph and our new code is built from optimal coloring of the constant weight single deletion graphs.
We showed that for multiple deletions, the best possible colorings are not guaranteed to produce codes meeting the Levenshtein upper bound.
If a coloring contains a color class that meet this upper bound, that class must be much larger than the average size of the classes in the coloring.

\appendices
\section{Counting Superstrings}
\label{section:counting-superstrings}
Let $x \in [2]^n$ and $y \in I_s(x)$ and consider a specific set of $s$ insertions that create $y$ from $x$.
Suppose $b$ is a new symbol and it is inserted immediately before $x_i$.
If $b = x_i$, then we can produce the same superstring by instead inserting $b$ immediately after $x_i$.
Consequently, to produce any supersting it is sufficient to use only two types of insertions:
insertions of the complement of $x_i$ before $x_i$, and arbitrary insertions at the end of $x$.
We would like to keep track of how many of our insertions are ones and how many are zeros.
New zeros can be inserted before existing ones in $x$ and at the end of $x$.
New ones can be inserted before existing zeros in $x$ and at the end of $x$.

We can make these ideas precise with the following bijection. 

\begin{lemma}
\label{lemma:bijection}
For each $x \in [2]^n_k$, there is a bijection between $I_{(s,r)}(x)$ and
\begin{equation*}
\bigcup_{a=0}^{s-r} \bigcup_{b=0}^r \left([2]^{n-k+b-1}_b \times [1] \right) \times \left([2]^{k+a-1}_a \times [1] \right) \times [2]^{s-a-b}_{r-b}.
\end{equation*}
\end{lemma}
\begin{IEEEproof}
We will refer to the latter set as encodings of the insertions that produce $y$ from $x$ and denote the set as $J_{(s,r),(n+s,k+r)}$.
We will describe the bijection explicitly as an encoding function from $I_{(s,r)}(x)$ to $J_{(s,r),(n+s,k+r)}$ and an inverse decoding function.

To describe these algorithms, we need a few simple string operations.
If a string is nonempty, it has a head that is a symbol and a tail that is another string.
We write the empty string as $\epsilon$.
We use a colon to indicate string concatenation.

First, we describe the encoding function.
\begin{algorithm}[H]
\caption{Encoding $y \in I_{(s,r)}(x)$ as $z \in J_{(s,r)(n+s,k+r)}$}
\begin{algorithmic}
\Procedure{Encode}{$x,y$}
\State $(z_0,z_1) \gets (\epsilon,\epsilon)$
\While{$x \neq \epsilon$}
\State $(u,x) \gets (\textsc{Head}(x), \textsc{Tail}(x))$
\State $(v,y) \gets (\textsc{Head}(y), \textsc{Tail}(y))$
\While {$v \neq u$}
\State $z_u \gets z_u : 1$
\State $(v,y) \gets (\textsc{Head}(y), \textsc{Tail}(y))$
\EndWhile
\State $z_u \gets z_u : 0$
\EndWhile
\State $z_2 \gets y$
\State \Return $z$
\EndProcedure
\end{algorithmic}
\end{algorithm}
\textsc{Encode} consumes symbols from $y$ until it finds one that matches the head of $x$.
It add a one to the output for each mismatch and adds a zero when it finally finds a match.
Which output it uses depends on the current head of $x$
When $x$ runs out of symbols, any remaining symbols in $y$ become the third part of the output.

The first term of the product, $z_0$ specifies how many new ones to insert before each existing zero.
The number of zeros in $z_0$ is equal to the number of zeros in $x$ and the last symbol of $z_0$ is always zero.
The total number of ones inserted this way is $b$, so $z_0 \in \left([2]^{n-k+b-1}_b \times [1] \right)$ for some $b$.
The second term of the product specifies how many new zeros to insert before each existing one.
The total number of zeros inserted this way is $a$, so $z_1 \in \left([2]^{k+a-1}_a \times [1] \right)$ for some $a$.
The third term specifies the insertions at the end of the string.
There must be $s-r-a$ zeros and $r-b$ ones inserted there, so $z_2 \in [2]^{s-a-b}_{r-b}$.

Now we describe the decoding function:
\begin{algorithm}[H]
\caption{Decoding $y \in I_{(s,r)}(x)$ from $z \in J_{(s,r)(n+s,k+r)}$}
\begin{algorithmic}
\Procedure{Decode}{$x,z$}
\State $y \gets \epsilon$
\While{$x \neq \epsilon$}
\State $(u,x) \gets (\textsc{Head}(x), \textsc{Tail}(x))$
\State $(w,z_u) \gets (\textsc{Head}(z_u), \textsc{Tail}(z_u))$
\While {$w = 1$}
\State $y \gets y : \overline{u}$
\State $(w,z_u) \gets (\textsc{Head}(z_u), \textsc{Tail}(z_u))$
\EndWhile
\State $y \gets y : u$
\EndWhile
\State $y \gets y : z_2$
\State \Return $y$
\EndProcedure
\end{algorithmic}
\end{algorithm}
The head of $x$ determines whether \textsc{Decode} inspects $z_0$ or $z_1$.
\textsc{Decode} adds the complement of the head of $x$ to the output for each one in $z_b$.
When it finds a zero, it outputs the head of $x$ and advances.
When $g$ reaches the end of $x$, it adds $z_2$ to the output.

It is easy to see that if $y \in I_{(s,r)}(x)$ and $\textsc{Encode}(x,y) = z$, then $\textsc{Decode}(x,z) = y$.
Fig.~\ref{figure:alg} illustrates an example execution of each algorithm.

\begin{figure}
\subfigure[ ]{
\begin{tabular}{|r|r|l|l|l|}
\hline
$x$ & $y$ & $z_0$ & $z_1$ & $z_2$ \\
\hline
0110001 & 001001010101101 & & & \\
110001 & 01001010101101 & 0 & & \\
10001 & 001010101101 & 0 & 10 & \\
0001 & 010101101 & 0 & 10110 & \\
001 & 10101101 & 00 & 10110 & \\
01 & 101101 & 0010 & 10110 & \\
1 & 1101 & 001010 & 10110 & \\
 & 101 & 001010 & 101100 & \\
 & & 001010 & 101100 & 101\\
\hline 
\end{tabular}\label{subfig:encode}}
\subfigure[]{\begin{tabular}{|r|r|r|r|l|}
\hline
$x$ & $z_0$ & $z_1$ & $z_2$ & $y$ \\
\hline
0110001 & 001010 & 101100 & 101 & \\
110001 & 01010 & 101100 & 101 & 0 \\
10001 & 01010 & 1100 & 101 & 001 \\
0001 & 01010 & 0 & 101 & 001001 \\
001 & 1010 & 0 & 101 & 0010010 \\
01 & 10 & 0 & 101 & 001001010 \\
1 &  & 0 & 101 & 00100101010 \\
 &  &  & 101 & 001001010101 \\
 &  &  & & 001001010101101 \\

\hline
\end{tabular}\label{subfig:decode}}
\caption{
The table \ref{subfig:encode} illustrates the computation of \textsc{Encode}$(x,y)$ and table \ref{subfig:decode} illustrates the computation of \textsc{Decode}$(x,z)$.
In each case, there is a row for each iteration of the outer while loop.
}
\label{figure:alg}
\end{figure}
\end{IEEEproof}

We will use a few well known combinitorial results in following lemma and the lemma in the next appendix.

Recall that Vandermonde's identity is
\begin{equation}
\label{eq:vandermonde}
\binom{a+b}{c} = \sum_{i=0}^c \binom{a}{i} \binom{b}{c-i}.
\end{equation}
This bijection correpsonding to this identity splits a string of length $a+b$ into a string of length $a$ and a string of length $b$.
The sum is over all possible distributions of $c$ ones in the original string between the new strings.

The number of multisets with $n$ possible unique elements and $k$ elements is $\binom{n+k-1}{k}$.
Such a multiset can be represented as a string of $n-1$ zeros and $k$ ones.
Each one corresponds to an element and the zero mark the boundaries between different types of elements.
A version of Vandermonde's identity related to multiset counting is
\begin{equation}
\label{eq:vandermonde-multi}
\binom{a+b+c-1}{c} = \sum_{i=0}^c \binom{a+i-1}{i} \binom{b+c-i-1}{c-i}.
\end{equation}
This decomposes a multiset with $a+b$ possible unique elements into a multiset with $a$ possible unique elements and a multiset with $b$ possible unique elements.
This also corresponds to breaking a string with $a+b-1$ zeros at the location of its $a$th zero, so $a-1$ zeros are in the first fragment and $b-1$ are in the second fragment.

\newtheorem*{L2}{Lemma \ref{lemma:layer-insertion-number}}
\begin{L2}
For all $n,k,s,r \in \mathbb{N}$ with $0 \leq r \leq s \leq n$ and $0 \leq k \leq n$, and all $x \in [2]^{n-s}_{k-r}$, the number of superstrings of $x$ with length $n$ and weight $k$ satisfies $|I_{(s,r)}(x)| =  \sum_{i=0}^{\min(r,s-r)} \binom{k+s-2r}{s-r-i} \binom{n-k-s+2r}{r-i}$.
\end{L2}
\begin{IEEEproof}
From Lemma~\ref{lemma:bijection}, for all $x\in [2]^n_k$, $|I_{(s,r)}(x)| = |J_{(s,r)(n+s,k+r)}|$.
This value is
\begin{equation*}
 \sum_{a=0}^{s-r} \sum_{b=0}^r \binom{n-k+b-1}{b} \binom{k+a-1}{a} \binom{s-a-b}{r-b}.
\end{equation*}
From Vandermonde's identity, \eqref{eq:vandermonde}, we have
\begin{IEEEeqnarray*}{Ccl}
 \binom{s-a-b}{r-b} &=& \sum_{c=0}^{r-b} \binom{s-r-a}{c}\binom{r-b}{r-b-c}, \\
 &=& \sum_{c=0}^{\min(r,s-r)} \binom{s-r-a}{s-r-a-c}\binom{r-b}{r-b-c}.
\end{IEEEeqnarray*}
Substituting this into the expression for $I_{(s,r),(n+s,k+r)}$ and exchanging the order of the sums yields
\begin{IEEEeqnarray*}{Ccl}
&\sum_{c=0}^{\min(r,s-r)} & \left( \sum_{a=0}^{s-r} \binom{k+a-1}{a} \binom{s-r-a}{s-r-a-c} \right. \\
&& \left. \sum_{b=0}^r \binom{n-k+b-1}{b} \binom{r-b}{r-b-c} \right). \\
\end{IEEEeqnarray*}
The multiset variant of Vandermonde's identity, \eqref{eq:vandermonde-multi}, eliminates the sums over $a$ and $b$ giving
\begin{equation*}
\sum_{c=0}^{\min(r,s-r)} \binom{k+s-r}{s-r-c}\binom{n-k+r}{r-c}.
\end{equation*}
Substituting $n-s$ for $n$ and $k-r$ for $k$ yields the claimed result.
\end{IEEEproof}

\section{Proof of Lemma \ref{lemma:sum-bound}}
\label{section:proof-of-sum-bound}
\newtheorem*{L3}{Lemma \ref{lemma:sum-bound}}
\begin{L3}
For all $s \in \mathbb{N}$,
\begin{equation*}
f_s(p) = \sum_{0 \leq r \leq s} \binom{s}{r}^2 p^{s-r}(1-p)^r,
\end{equation*}
is maximized at $p = 1/2$, so for all $p$, $f_s(p) \leq 2^{-s} \binom{2s}{s}$.
\end{L3}
\begin{IEEEproof}
To obtain the required upper bound, we express $f_s(p)$ as $\sum_{i \geq 0} a_i p^i(1-p)^i$ where all $a_i$ are nonnegative.

Starting from Vandermonde's identity, \eqref{eq:vandermonde}, we can derive 
\begin{IEEEeqnarray*}{rCl}
\binom{s}{r}^2 & = & \binom{s}{r}\sum_i \binom{r}{i}\binom{s-r}{s-r-i}, \\
& = & \sum_i \binom{s}{i,i,r-i,s-r-i}.
\end{IEEEeqnarray*}
Two of the four parts of the multinomial coefficient involve $r$.
Isolating these yields 
\begin{equation}
\label{equation}
\binom{s}{r}^2 = \sum_i \binom{s}{i,i,s-2i} \binom{s-2i}{r-i},
\end{equation}
which will allow us to perform the desired change of basis.
Crucially, the first binomial coefficient does not depend on $r$.
Applying \eqref{equation} to $f(p)$ yields
\begin{IEEEeqnarray*}{rCl}
f_s(p) & = & \sum_{0 \leq r \leq s}  p^{s-r}(1-p)^r \sum_i \binom{s}{i,i,s-2i} \binom{s-2i}{r-i}, \\
& = & \sum_i \binom{s}{i,i,s-2i} p^i(1-p)^i \times \\
&& \quad \quad \sum_{0 \leq r \leq s}  p^{s-r-i}(1-p)^{r-i} \binom{s-2i}{r-i}, \\
& = & \sum_i \binom{s}{i,i,s-2i} p^i(1-p)^i .
\end{IEEEeqnarray*}
The binomial theorem reduced the internal sum to 1.
Applying $p(1-p) \leq 1/4$ yields
\begin{IEEEeqnarray*}{rCl}
f_s(p) & \leq & \sum_i \binom{s}{i,i,s-2i} 2^{-2i}, \\
& = & 2^{-s} \sum_i \binom{s}{i,i,s-2i} \sum_j \binom{s-2i}{j-i}, \\
& = & 2^{-s} \sum_j \binom{s}{j}^2, \\
& = & 2^{-s} \binom{2s}{s}.
\end{IEEEeqnarray*}
We undo the change of basis by expanding $2^{s-2i}$ with the binomial theorem, reordering the sums, and applying \eqref{equation}.
Finally Vandermonde's identity eliminates the sum.
\end{IEEEproof}

\section{Induced Subgraphs and Graph Perfectness}
A graph $G$ is perfect if and only if for each induced subgraph $H$, $\omega(H) = \chi(H)$ \cite{west_introduction_2001}.
This is a hereditary property.
A graph is perfect if and only if all of its induced subgraphs are perfect.

\label{section:induced}
\begin{lemma}
\label{lemma:reduce-n}
$L_{s,n}$ is an induced subgraph of $L_{s,n+1}$.
\end{lemma}
\begin{IEEEproof}
Take the vertices of $L_{s,n+1}$ corresponding to the strings that begin with 0.
\end{IEEEproof}

\begin{lemma}
\label{lemma:cycle}
Let $C_{n}$ be the cyclic graph with $n$ vertices. For all $n \in \mathbb{N}$ with $n \geq 3$, $C_n$ is an induced subgraph of $L_{s,(n-2)s+1}$.
\end{lemma}
\begin{IEEEproof}
We will pick strings $x_i,y,z \in [2]^{(n-2)s+1}$ for $0 \leq i \leq n-3$.
For all $0 \leq i \leq n-3$, $x_i = 0^{si}1^{s+1}0^{s(n-3-i)}$, $y = 10^{(n-2)s}$ and $z = 0^{(n-2)s}1$.
Then for $0 \leq i \leq n-4$, $d_L(x_i,x_{i+1}) = s$, $d_L(x_0,y) = s$, $d_L(x_{n-3},z) = s$, $d_L(y,z) = 1$, and all other distances are greater than $s$.
\end{IEEEproof}
As an example, for $s=1$ and $n=5$ we pick 1100,0110,0011,0001, and 1000.

\newtheorem*{T1}{Theorem \ref{theorem:imperfect}}
\begin{T1}
For all $s,n \in \mathbb{N}$ with $s \geq 1$ and $n \geq 3s+1$, $L_{s,n}$ is not a perfect graph.
\end{T1}
\begin{IEEEproof}
By Lemma~\ref{lemma:reduce-n}, $L_{s,3s+1}$ is an induced subgraph of $L_{s,n}$.
By Lemma~\ref{lemma:cycle}, the five cycle is an induced subgraph of $L_{s,3s+1}$.
Odd cycles with at least five vertices are not perfect because a proper coloring requires three colors even though their largest clique contains only two vertices.
\end{IEEEproof}

\bibliographystyle{IEEEtran}
\bibliography{IEEEabrv,cullina}

\end{document}